\documentclass[floatfix,secnumarabic,amssymb,nobibnotes,nofootinbib,preprint,aps,pra, superscriptaddress,showpacs]{revtex4-2}
\usepackage{amsmath}
\usepackage{amsfonts}
\usepackage{amssymb}
\usepackage{graphicx}
\usepackage{units}
\usepackage{color}
\usepackage{xspace}
\usepackage[normalem]{ulem}
\usepackage{natbib}
\usepackage{textcomp}
\DeclareUnicodeCharacter{2009}{\,}
\DeclareUnicodeCharacter{2212}{-}

\usepackage[symbol*]{footmisc}

\DefineFNsymbolsTM{otherfnsymbols}{%
  \textdagger    \dagger
  \textdaggerdbl \ddagger
  \textasteriskcentered *
  \textbardbl    \|%
  \textparagraph \mathparagraph
}%

\setfnsymbol{otherfnsymbols}

\begin{document}

\title{Negative Capacitance for Stabilizing Logic State in Tunnel Field-Effect Transistor}

\author{Koushik Dey}
\email{koushikdey664@gmail.com}
\affiliation{School of Physical Sciences, Indian Association for the Cultivation of Science, 2A $\&$ B
Raja S. C. Mullick Road, Jadavpur, Kolkata - 700032, India
}

\author{Bikash Das}
% bikashcgr1@gmail.com
\affiliation{School of Physical Sciences, Indian Association for the Cultivation of Science, 2A $\&$ B
Raja S. C. Mullick Road, Jadavpur, Kolkata - 700032, India
}

\author{Pabitra Kumar Hazra}
% pabitrahazra7495@gmail.com 
\affiliation{School of Physical Sciences, Indian Association for the Cultivation of Science, 2A $\&$ B
Raja S. C. Mullick Road, Jadavpur, Kolkata - 700032, India
}

\author{Tanima Kundu}
% tanima.kundu96@gmail.com
\affiliation{School of Physical Sciences, Indian Association for the Cultivation of Science, 2A $\&$ B
Raja S. C. Mullick Road, Jadavpur, Kolkata - 700032, India
}

\author{Sanjib Naskar}
% mssn@iacs.res.in 
\affiliation{Central Scientific Services, Indian Association for the Cultivation of Science, 2A $\&$ B
Raja S. C. Mullick Road, Jadavpur, Kolkata 700032, India
}

\author{Soumik Das}
% dassoumik555@gmail.com 
\affiliation{School of Physical Sciences, Indian Association for the Cultivation of Science, 2A $\&$ B
Raja S. C. Mullick Road, Jadavpur, Kolkata - 700032, India
}
\author{Sujan Maity}
% sujanmaity1994@gmail.com 
\affiliation{School of Physical Sciences, Indian Association for the Cultivation of Science, 2A $\&$ B
Raja S. C. Mullick Road, Jadavpur, Kolkata - 700032, India
}
\author{Poulomi Maji}
% poulomimaji8@gmail.com
\affiliation{School of Physical Sciences, Indian Association for the Cultivation of Science, 2A $\&$ B
Raja S. C. Mullick Road, Jadavpur, Kolkata - 700032, India
}
\author{Bipul Karmakar}
% spsbk2647@iacs.res.in
\affiliation{School of Physical Sciences, Indian Association for the Cultivation of Science, 2A $\&$ B
Raja S. C. Mullick Road, Jadavpur, Kolkata - 700032, India
}

\author{Rahul Paramanik}
% rahulparamanik456@gmail.com
\affiliation{School of Physical Sciences, Indian Association for the Cultivation of Science, 2A $\&$ B
Raja S. C. Mullick Road, Jadavpur, Kolkata - 700032, India
}

\author{Subhadeep Datta}
\email{sspsdd@iacs.res.in}
\affiliation{School of Physical Sciences, Indian Association for the Cultivation of Science, 2A $\&$ B
Raja S. C. Mullick Road, Jadavpur, Kolkata - 700032, India
}

%========================================================================================================

\begin{abstract}

Ferroelectric negative capacitance transistors, or Fe-NCFETs, are promising device architecture for achieving improved performance in terms of hysteresis, on-off ratio and power consumption. The study investigates the influence of negetive capacitance (NC) on the transfer characteristics of van der Waals Field-Effect Transistors (vdW FETs) below and above a critical voltage (V$_{th}$) on the hetero phase of CuInP$_{2}$S$_{6}$ (CIPS) gate ferroelectric. Notably, a less pronounced NC resulting from the spatial distribution of the ferroelectric (FE) and paraelectric (PE) phases plays a crucial role in stabilizing  of n-channel conductance by dual gate modulation. This results into the emergence of a non-volatile logic state, between the two binary states typical of conventional Tunnel Field-Effect Transistors (TFETs). Concerned study proposed negative capacitance tunnel field-effect transistors (NC-TFETs) based on ferroionic crystals as promising devices for generating a stable logic state below the coercive voltage. In addition, tunneling and voltage pinning effects play key role for enhancement of transistor's on-off ratio.

keywords: Negetive capacitance, Field effect transistor, 2D ferroelectric, TMDs, Tunnel FET, logic state.

\end{abstract}

\maketitle

\section{Introduction}  

Tunnel field-effect transistors (TFETs) have garnered a lot of attention due to their ability to achieve a steep subthreshold swing (SS) with band-to-band tunneling (BTBT) of carriers \cite{BTBT,Lee_M}. However, the on-current of conventional TFETs is too low to be a viable substitute for metal oxide semiconductor field-effect transistors (MOSFET). Integrating ferroelectric (FE) materials with TFETs has been proved to be a useful way to overcome this important technological issues \cite{TFET_SSS}. Ferroelectric materials exhibit negetive capacitance (NC) region which in general highly unstable region and cannot be realized through experiments \cite{NC_Salahuddin}. However, this NC effect can be stabilized by placing suitable dielectric in series with a FE capacitor \cite{Saeidi_IEEE,Khan_Natmat}. Furthermore, a MOS transistor's gate capacitance can be increased by integrating a ferroelectric capacitor into the gate stack, particularly when the transistor is functioning in the NC region \cite{Park_SR}. This series architecture can act as a step up voltage transformer, resulting in a sudden rise in the internal voltage which is proportional to the polarization of ferroelectric materials. Consequently, the probability of BTBT increases and the drain current enhances \cite{step_upT}.

Nowadays, p-type metal oxide (MO) materials like SnO, CuO, and Cu$_{2}$O have attracted considerable attention due to their distinctive properties and potential applications in tunnel field-effect transistors (TFETs). Moreover, to facilitate monolithic three-dimensional (M3D) integration of high-performance logic, there is a strong demand for high-mobility p-type oxide semiconductor thin films. These thin films can be utilized in the fabrication of back-end-of-line (BEOL) compatible complementary metal-oxide-semiconductor (CMOS) circuits \cite{Bae_ultrathin}. Among these, atomic layer deposited Cu-based oxide CMOS devices are promising for high performance in terms of high mobility, low off current etc \cite{Maeng_Atomic,Shen_atomic,Lee_Cu2O,Shih_new}. Thus, these are attracting for real-world applications i.e, photonic, spectroscopic, photoelectronics and energy harvesting purposes \cite{Gyawali_photocarrier,Huang_GaAs,Moham_Cu,Gyawali_size}. 
 
On the contrary, wide-bandgap metal-oxide n-type semiconductors capable of sustaining a robust p-type inversion layer when paired with a high-dielectric-constant (HZO) barrier and sourced with a heterogeneous p-type material. The utility of this inversion layer as for controlling element in a unique tunnelling junction transistor result into remarkable characteristics, such as high current, power, and transconductance densities \cite{Shoute_sustained}. Here, we integrate a wide bandgap n-type transition metal dichalcogenides (TMDs) WS$_{2}$ paired with a ferroelectric material for TFET device architecture.

Transistors incorporating newly developed two-dimensional (2D) van der Waals (vdW) ferroelectric materials outperform conventional complementary metal-oxide-semiconductor FETs (CMOSFETs), exhibiting reduced power consumption, increased switching speed, and enhanced stability \cite{MoS2_CIPS,Liao_ACSnano}. The potential of vdWs FeFETs to surpass the limits of Moore's law opens up new possibilities for exploration and utilization in nanoelectronics, particularly in nonvolatile memory and neuromorphic computing applications \cite{Venema,Duhan}. The atomically thin channels of 2D semiconductors, like transition metal dichalcogenides (TMDs), provide ideal electrostatic control to improve immunity to short channel effects \cite{short_channel}. For this reason, they have been thoroughly investigated as channel materials for future electronic device applications. Negetive capacitance field effect transistors (NCFETs) have been documented extensively using 2D TMDs as the channel material and ferroelectric materials such as hafnium zirconium oxide (HZO) as the ferroelectric gate \cite{Chen_HZO}. In comparison to bulk ferroelectrics, layered ferroelectrics with atomically smooth surfaces may provide superior performance and reliability for NC-FETs by minimizing the impact of dangling bonds and charged impurities that induce interface traps \cite{Zubko}. 

CuInP$_{2}$S$_{6}$ (CIPS) is such a ferroelectric 2D materials which exhibits switchable polarization down to 4 nm at room temperature \cite{Io_4} and enables interface integration for constructing vdW heterostructure-based functional memory and logic devices like NC-FETs, ferroelectric tunnel junctions (FTJs) etc \cite{MoS2_CIPS,FTJ}. Moreover, CIPS is recognized for its significant electric-field controllable macroscopic ionic conductivity \cite{Jiang_manipulation}. The interplay between this conductivity and ferroelectricity give rises to ferroionic states that can be useful for getting novel device performance \cite{Anisotropic_NL}. It's worth mentioning that in cases of copper deficiency in CIPS, the system experiences a chemical phase separation, leading to the formation of a paraelectric (PE) In$_{4/3}$P$_{2}$S$_{6}$ (IPS) phase alongside ferroelectric CIPS phase \cite{Susner_spinodal}. The formation of FE-PE hetero-structure by spatial random distributions may results into performance variations when integrated on a TMD channel and may open up new possibilities for getting different logical devices \cite{NC_reduction}. 

To generate desirable and stable logical device, we have implemented two strategies i.e. NC modulation and the effect of long-range Cu ion migration in a vertical CIPS-IPS/WS$_{2}$ heterostructure. Here, we report the influence of NC on the transfer characteristics of vdW FETs below and above a critical voltage (V$_{th}$) on the CIPS-IPS gate. This study reveals the important role of a less pronounced NC resulting from the spatial distribution of the FE and PE phases for the stabilization of transconductance below the V$_{th}$, leading to the establishment of a nonvolatile logic state between the two binary states typical of conventional TFETs. 

%%%%%%%%%%%%%%%%%%%%%%%%%%%%%%%%%%%%
\section{Experimental details}
\label{subsec:exp}
%%%%%%%%%%%%%%%%%%%%%%%%%%%%%%%%%%%%

{\bfseries Synthesis and Characterization:} CIPS single crystals were synthesized \textit{via} standard chemical vapour transport (CVT) method by utilizing two zone tube furnace. Cu (99.95\%, Alfa Aesar), In (99.95\%, Alfa Aesar), P (99.95\%, Alfa Aesar) and S (99.99\%, Alfa Aesar) powders were mixed thoroughly in an atomic ratio 1:1:2:6. 5 mg/cc I$_{2}$ powder was used as transporting agent. The powders were thoroughly mixed and loaded into a quartz tube (25 cm) followed by evacuating the tube under 10$^{-5}$ Torr vacuum, sealed and then placed in a two-zone furnace. The temperature of the source zone was set to 750$^{\circ}$C, and the growth zone to 650$^{\circ}$C. After holding 7 days (168 hrs) at this temperature, the furnace was allowed to cool slowly to room temperature. CIPS single crystals were extracted by cleaving the chunk obtained inside the quartz tube. WS$_{2}$ single crystals were grown from starting material WS$_{2}$ micro-powder (99.95 \% Sigma-Aldrich) by sealing the powders with I$_{2}$ in a quartz tube, followed by evacuation. The tube was kept in a horizontal two-zone tube furnace for seven days. The growth zone's temperature was set at 840$^{\circ}$C, while the source zone's temperature was set to 920$^{\circ}$C. The as-grown WS$_{2}$ single crystals were extracted from the tube's cold regime.

X-ray diffraction (XRD) measurements were carried out using Bruker diffractometer using Cu K$_{\alpha}$ source on single crystal and on fine powders. Scanning electron microscope (SEM) and energy-dispersive spectroscopic (EDS) study were also performed in a JEOL JSM-6010LA. Polarization electric field (PE) hysteresis loops were recorded at room temperature on a thin crystal by utilizing Radiant precision ferroelectric tester. The Raman measurements were performed with a 532 nm laser in the backscattering geometry using a Horiba T64000 single spectrometer with high-resolution grating and a CCD detector. Temperature dependent dielectric measurements were carried out on a wet cryostat (OXFORD Optistat CF) utilizing Zurich Instruments MFIA impedance analyzer and CRYOCON (22c) temperature controller. Capacitance under dc biases on flakes were recorded with the same impedance analyzer at room temperature. Two probe I-V measurements were carried out on thin flakes by utilizing 2601B source measure units (SMU). A commercial atomic force microscope (Asylum Research MFP-3D) equipped with Ti/Ir-coated Si cantilever tip was used to detect piezoresponse on thin crystal flakes transferred onto Indium Tin oxide (ITO) coated glass substrate by utilizing dual ac resonance tracking piezoresponse force microscopy (DART-PFM) mode. Several distinct random spots were chosen to measure the local piezoresponse hysteresis loops by varying tip bias voltage on flakes. 

{\bfseries Device Fabrication and electrical characterization:} After micromechanical exfoliation of WS$_{2}$ crystals using standard scotch tape, WS$_{2}$ flakes were transferred onto a clean SiO$_2$ (285 nm)/Si substrate. Suitable flakes were located by utilizing optical microscope (OLYMPUS BX53M) by comparing the flake's transparency. Thin flakes of CIPS-IPS crystal was exfoliated on PDMS gelpak and were integrated on top of each selected WS$_{2}$ flakes by utilizing a micromanipulator. After that, optical lithography was employed to create the devices. Using a thermal evaporator, metal contacts were created with 10 nm Cr followed by 90 nm Au. Next, FET measurements were performed by utilizing SIGLENT's SDG7000A arbitrary waveform generator, Keithley 2450 and 2601B (SMU).  

%%%%%%%%%%%%%%%%%%%%%%%%%%%%%%%%%%%%
\section{Results and discussion}
\label{sec:results_discussions}
%%%%%%%%%%%%%%%%%%%%%%%%%%%%%%%%%%%%  

\begin{figure}
\centerline{\includegraphics[scale=0.35, clip]{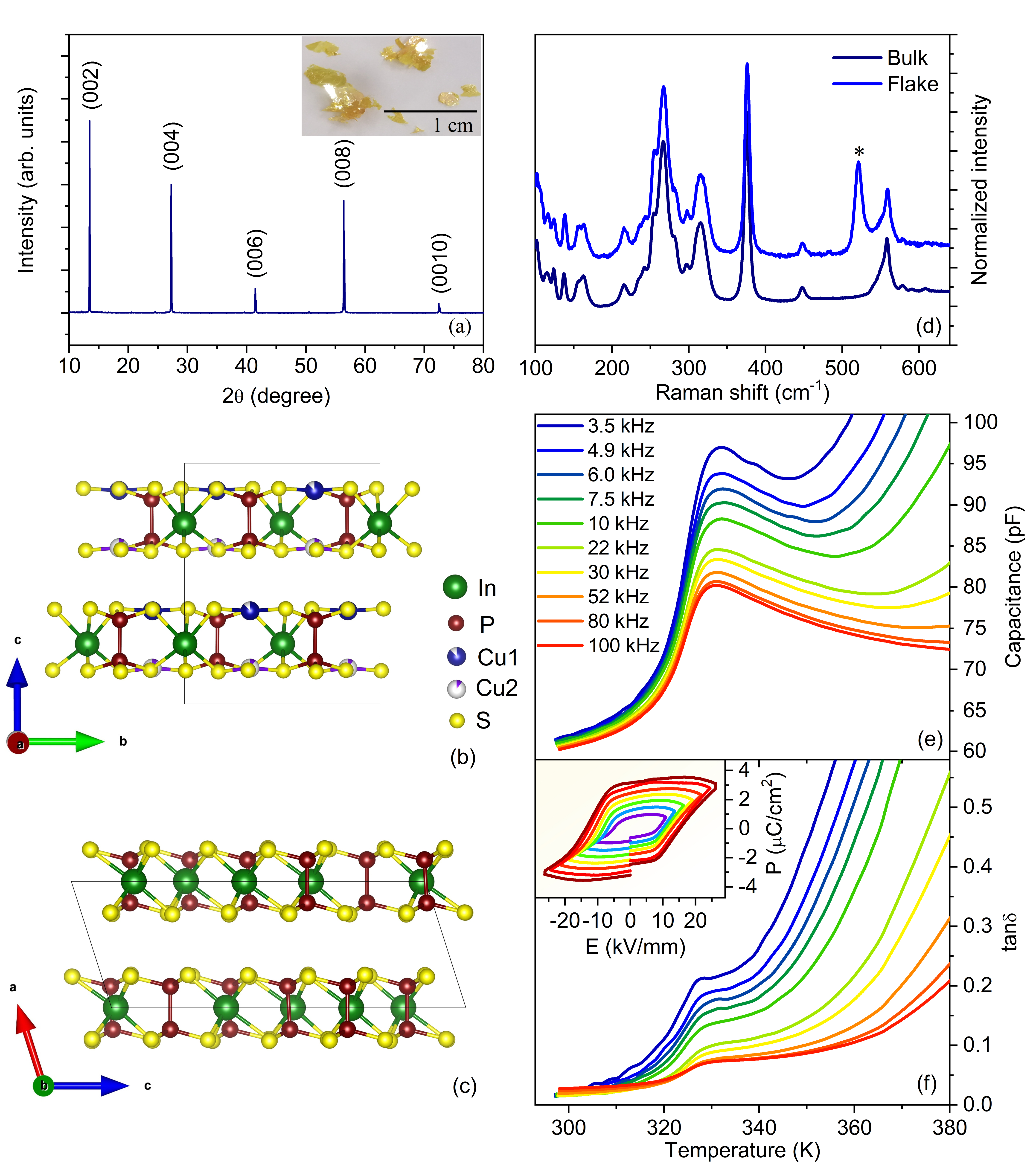}}
\caption{(a) X-ray diffraction pattern of ab crystalline plane of the single crystal. Crystal structure of (b) CIPS \& (c) IPS with vdW gap between the layers. The inset shows the as grown single crystals. (b) Room temperature Raman spectra of bulk crystal and thin flake. The starred mode corresponds to the characteristic mode of the Si wafer. Temperature dependence of (c)capacitance and (d) dielectric loss of the crystal at different frequencies. Frequency independent maxima $\sim$328 K in (c) \& (d) suggests a structural phase transition occurs at that temperature. The inset in (d) shows the trend of P-E loops up to various maximum electric fields at the same frequency.
\label{Figure1}}
\end{figure}

The as-grown single crystal exhibits transparent homogeneous colored thin sheets (see the inset of Figure \ref{Figure1}a). The quality of the crystal is examined through X-ray diffraction (XRD) of \textit{ab}-plane orientation. All diffraction peaks can be successfully indexed to (00l) peaks. SEM energy-dispersive X-ray spectroscopy (EDS) examination was carried out to verify the atomic stoichiometry of this sample. The compositions derived from the EDS analyses, presented in Table S1 are averages from typically around four data collection regions (see Figure S1 in the Supporting Information). The average stoichiometry is found to be Cu deficient with composition Cu$_{0.46}$In$_{1.21}$P$_{2.39}$S$_{5.93}$. Further, the structure refinement of the X-ray diffraction (XRD) pattern for these ground crystals has confirmed the presence of two different phases with $\sim$63\% CIPS and $\sim$37\% IPS. The detailed structural information along with the Rietveld refined XRD pattern is presented in the Supporting Information (see Figure S2 and Table S2).    

The atomic structure of CIPS is characterized by Cu, In, and P–P triangular patterns occupying the octahedral spaces within the sulfur framework as shown in the Figure \ref{Figure1}b. Weak van der Waals (vdW) interactions facilitate the vertical stacking of layers to form bulk crystals. A complete unit cell comprises two adjacent monolayers, essential for properly characterizing the material's symmetry due to the exchange of Cu and P-P pairs between layers . Each layer of the monoclinic (space group \textit{Cc}) CIPS crystal is made up of S$_{6}$ octahedra that encircle P-P pairs or metal cations (Cu, In) \cite{Maisonneuve}. Together, the phosphorus atoms and the S$_{6}$ octahedra form a structural backbone consisting of [P$_{2}$S$_{6}$]$^{4−}$ anion groups that ionically link with Cu and In cations organized hexagonally (see Figure \ref{Figure1}b). CIPS is a collinear two-sublattice ferrielectric system which exhibits a Curie temperature (T$_{C}$) of approximately 315 K \cite{Simon}. In the ferrielectric phase below T$_{C}$, spontaneous polarization occurs with the polar axis perpendicular to the layer plane. This polarization results from off-center ordering in the Cu sublattice and the displacement of cations from centrosymmetric sites in the In sublattice \cite{Mai_PRB}. Notably, when Cu deficits are present in CIPS, the material spontaneously phases out into paraelectric In$_{4/3}$P$_{2}$S$_{6}$ (IPS) with centrosymmetric crystal structure (space group \textit{P$2_1$C}) \cite{Susner_spinodal} and forrielectric CIPS domains inside the same single crystal, sharing an anionic [P$_{2}$S$_{6}$]$^{4−}$ (see Figure \ref{Figure1}c).

To further confirm the presence of two different composition and to check the homogeneity of these two phases, we have collected Raman spectra on several regions of a bulk crystal and on exfoliated flakes with different thicknesses. The room temperature representative Raman spectrum of bulk crystal and flake are depicted in Figure \ref{Figure1}d. There are 27 number of modes which can be resolved in those Raman spectra, attributed to the presence of both CIPS and IPS sub-lattices within the material by comparing with the reported Raman spectra of pure CIPS and CIPS-IPS heterophase \cite{Raman_, Raman_CIPS}. The increased number of modes in our Raman spectra is consistent with the CIPS-IPS Raman spectra of earlier reports. Interestingly, the Raman spectrum of the flake agrees well with that of bulk crystal which assures chemically homogeneous distribution of CIPS and IPS phases down to few layer flakes. 

Above 100 cm$^{-1}$, four different modes are observed in the anion translation region, ranging from 100 to 140 cm$^{-1}$. Among these modes, two (at $\sim$100 cm$^{-1}$ and $\sim$114 cm$^{-1}$) are attributed to CIPS, while the other two (127 cm$^{-1}$ and 140 cm$^{-1}$) are assigned to IPS \cite{PRM}. Similarly, in the range of 200 to 300 cm$^{-1}$, the two most intense peaks are observed at $\sim$255 and $\sim$270 cm$^{-1}$, corresponding to IPS and CIPS, respectively. Generally, mode at $\sim$300 cm$^{-1}$ can not be seen in the pure CIPS, which is attributed to anion deformation in IPS \cite{Raman_CIPS}. Whereas, the made at $\sim$ 320 cm$^{-1}$ is assigned to anion deformation in CIPS. Raman modes at $\sim$580 and $\sim$610 cm$^{-1}$ do not appear in the CIPS Raman spectrum and are considered as the anion stretching modes of IPS \cite{PRM}.

The interplay between the two sub-lattices in the CIPS-IPS heterostructure causes the IPS domains to exert a chemical pressure over the CIPS domains, increasing the overall T$_{C}$ to approximately 320-335 K for substantially Cu-deficient CIPS \cite{Checa}. 
Temperature-dependent dielectric results at selected frequencies are presented in Figure \ref{Figure1}(e-f). At low frequencies, dielectric losses rise with increasing temperature, resulting in an increase in the real part of the dielectric permittivity. This effect is primarily due to high conductivity \cite{Banys_die}. At higher frequencies, the real part of the dielectric permittivity aligns with the static permittivity, due to the significant decrement of tan$\delta$. It has been observed that Cu deficiencies or the addition of extra In ions significantly alter the ferroelectric phase transition temperature. From temperature-dependent dielectric measurements, a phase transition can be readily seen at around 328 K (maximum temperature of static dielectric permittivity), indicating a paraelectric to ferroelectric phase transition occurring at a higher temperature compared to pure CIPS.
%From temperature dependent dielectric measurements, a phase transition can be readily seen at $\sim$ 328 K temperature which confirms a paraelectric to ferroelectric phase transition occurs at an elevated temperature than that of pure CIPS. 
In order to confirm the ferroelectric properties of this material, polarization (P) vs out-of-plane electric field (E) data was measured on a thick flake. The presence of a distinct hysteresis loop (see the inset of Figure \ref{Figure1}f) is a direct evidence of ferroelectricity with coercive field (E$_{c}$) $\sim$16.4 kV/mm and remanent polarization (P$_{r}$) $\sim$3.42 $\mu$C/cm$^{2}$. We have recorded temperature dependent P-E loops on CIPS-IPS bulk crystals below and above the ferroelectric to paraelectric phase transition temperature, presented in the Supporting Information. Polarization loops and the corresponding current density loops at selected temperatures are shown in Figure S3(a-c). The polarization loops getting wider as we increased temperature. The changes in loop shape (\textit{i.e}, loop edges shift towards electric field axis with larger loop area) at higher fields and higher temperatures are due to the generation of larger leakage currents. From the current density plots, an abrupt increment of currents at higher fields (in FE phase) and almost linear rise of currents with fields (in PE phase) can be clearly seen. This indicates a strong coupling of ferroelectric polarization with ionic conduction in the CIPS-IPS heterophase. From temperature dependent E$_{c}$ (see the inset in Figure S3) it can be seen that CIPS-IPS completely loses its ferroelectric nature above 328 K.

\begin{figure}
\centerline{\includegraphics[width=\linewidth]{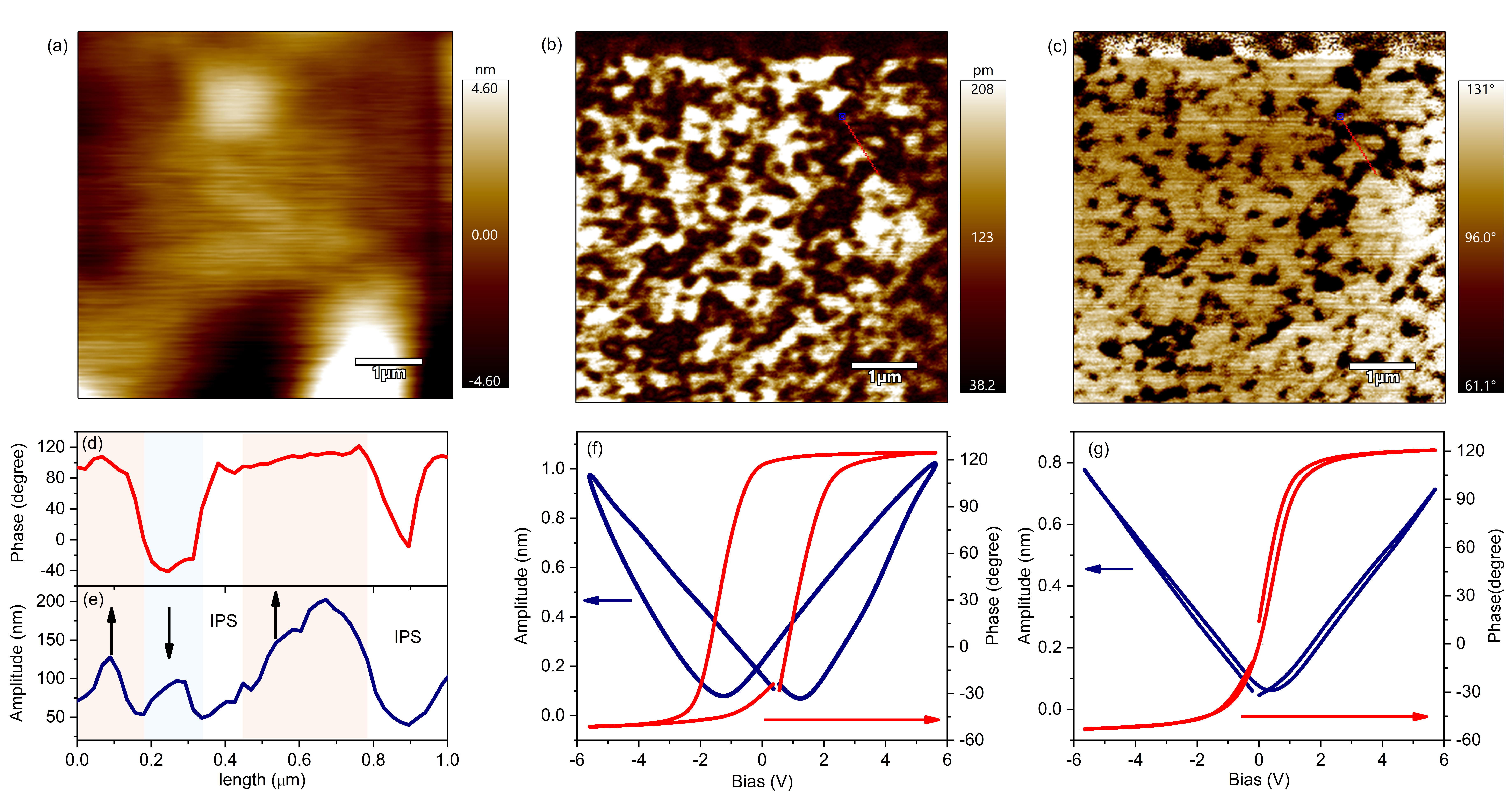}}
\caption{ (a) Topography, (b) PFM amplitude and (c) PFM phase images of the selected region of an exfoliated flake on ITO substrate. The line profile signals (red lines in (b) and (c)) of (d) PFM phase and (e) amplitude. A combination of these two signals separate out chemically different regions of CIPS-IPS self assembled hetero-structures. Arrow indicates polarization direction of domains of CIPS. Local PFM amplitude and phase hysteresis loops with dc bias on (f) CIPS and (g) IPS regions.    
\label{Figure2}}
\end{figure}

Next, we focus on piezoresponse force microscopy (PFM) results of CIPS-IPS flakes exfoliated on a metal substrate to investigate thin flake ferroelectric and piezoelectric properties. Figure \ref{Figure2}a illustrates the smooth topography of a representative CIPS-IPS flake. Fig. \ref{Figure2}b and \ref{Figure2}c represents PFM amplitude and phase map of the mentioned region. Indeed, the PFM amplitude serves as an indicator of the absolute magnitude of the local piezoelectric response. Simultaneously, the PFM phase provides information about the polarization direction within each individual domain. These measurements are valuable for characterizing the piezoelectric and ferroelectric properties of materials at the nanoscale \cite{Buragohain}. 

PFM measurements can distinguish between regions of different piezoresponse in the sample by closely examining the local piezoresponse of different regions within the flake. As illustrated in Figure \ref{Figure1}e-f the CIPS ferroelectric phase with upward and downward direction polarization domains are identified with higher \& lower PFM amplitude between those regions (arrows indicate polarization direction of the shaded regions). A clear 180$^{\circ}$C phase difference is observed between the upward- and downward-polarized ferroelectric domains within the CIPS phase. However, the IPS paraelectric phase shows negligible piezoresponse. This observation indicates CIPS and IPS regions spatially extends upto several nanometers. To further check piezoresponse in those mentioned regions we have recorded local PFM amplitude and phase hysteresis loops by varying dc bias, as shown in Figure \ref{Figure2}e-f. CIPS region exhibits a well defined strain and phase loop with a remanence, confirming ferroelectric nature of CIPS. Whereas, the IPS region shows a negligible strain and phase hysteresis loops similar like paraelectric phase.

\begin{figure}
\centerline{\includegraphics[scale=0.5, clip]{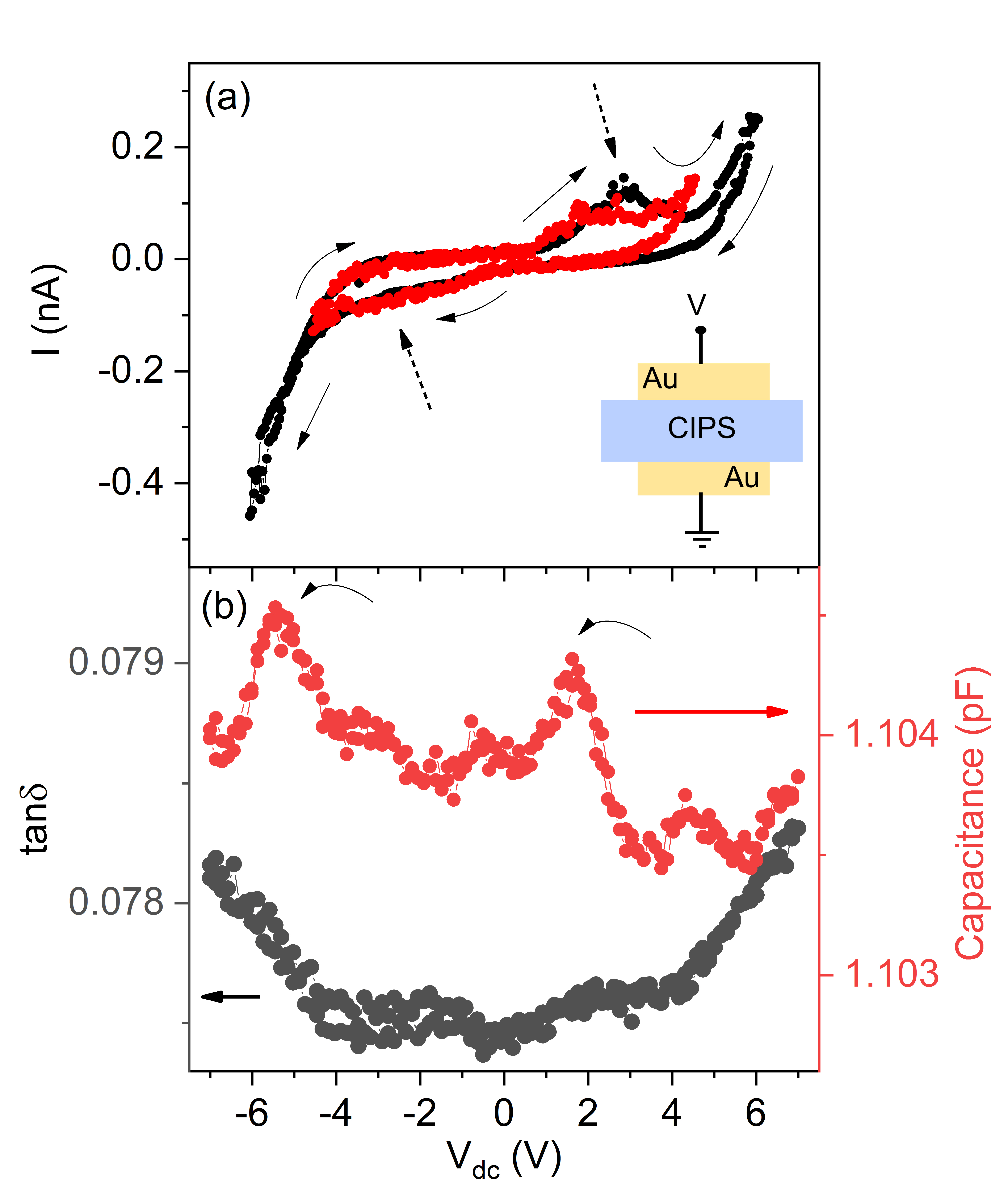}}
\caption{ (a) I-V curve of CIPS-IPS flake up to different maximum voltages on MIM device (schematic). Line arrows indicate the current path. Thick dashed arrow indicates two different humps where a broad current peaks appear due to polarization switching(b) Capacitance and dielectric loss \textit{vs.} dc bias. Two capacitence peak marked as curved arrows indicate polarization switching at those voltages.       
\label{Figure3}}
\end{figure}

To investigate the electric properties of CIPS-IPS flake, a vertical sandwiched Au/CIPS-IPS/Au device is fabricated [see inset of Figure \ref{Figure3}a]. I–V loops are measured at room temperature depicted in Fig. \ref{Figure3} (a). The cycled I–V curve reveals significant current hysteresis with a clockwise direction (shown by arrow). The magnitude of the hysteresis window strongly depends on the range of the maximum voltages. Notably, when examining the voltage sweep within the range of $\pm$ 4 volts (red), negligible current is observed. However, as the voltage range exceeds $\pm$ 5 V, the current starts to increase rapidly. The hump at around $\pm$ 3 volt (shown by dashed arrows) indicates current increment for polarization switching (nearby V$_{c}$), which typically drops above that for typical ferroelectrics. However, at higher voltages, a steep rise in current indicates the onset of long-range conduction. A sudden increment of dielectric loss (tan$\delta$) is also seen around the same voltage (see in Figure \ref{Figure3}b) which corroborates long range conduction at higher voltages.

The presence of ferroelectricity and paraelectricity in thin flakes can be obtained from the quasistatic capacitance-voltage (C-V) cmeasurements \cite{Choi}. C-V measurement involves superimposing a small amplitude AC voltage (to measure C) over a DC voltage (P). We have applied 500 mV peak-to-peak ac signal with 300 kHz frequency superimposed with a dc bias (varying upto $\pm$ 6 V) for the measurements. The two peaks in C-V characteristics (shown in Figure \ref{Figure3}b) associated with the ferroelectric switching phenomena. An imprint voltage of around 2 V is observed in the hysteresis loop of the C-V curve, indicating the presence of an internal electric field between electrodes. The internal electric field may generates due to the gradient in the strain distribution in the flake or a non-homogeneous spatial distribution of defects like Cu$^{+}$ vacancies \cite{Boni}. 

\begin{figure}
\centerline{\includegraphics[scale=0.25, clip]{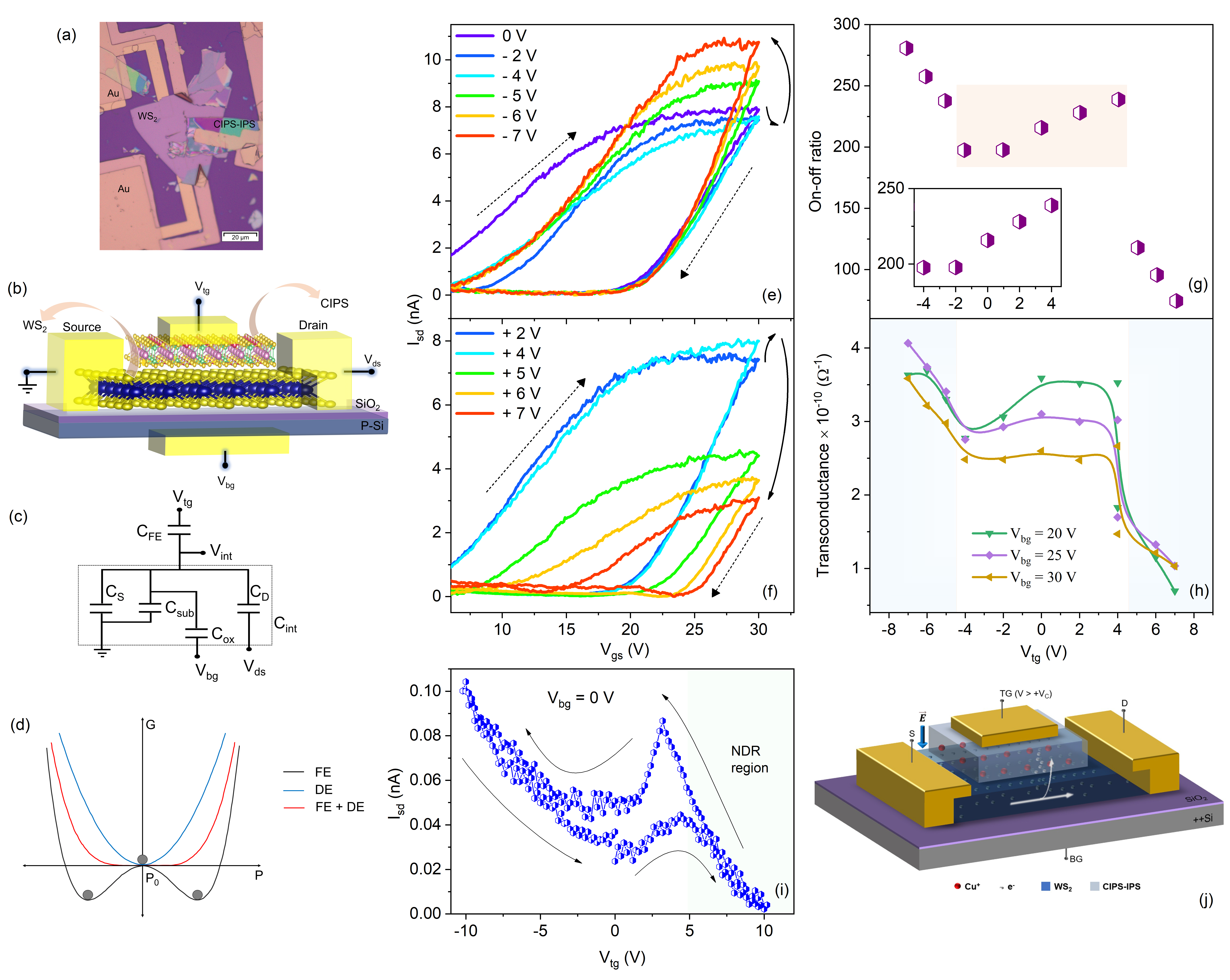}}
\caption{ (a) Optical micrograph of CIPS-IPS/WS$_{2}$ heterostructure integrated on SiO$_{2}$/Si substrate. (b)Schematic diagram of dual gate experiments on that heterostructure. (c) Equivalent capacitance model of NC-FET. (d) Gibbs free energy profiles at zero external electric field, plotted as a function of polarization, for ferroelectric (FE), dielectric (DE), and the combined system (FE+DE). For typical ferroelectrics, one of the two energy minima is occupied. However, when a dielectric is added in series with the ferroelectric, the energy profile of the system changes. A minimum is observed at P$_0$, where the ferroelectric resides in the negative capacitance state. I$_{ds}$-V$_{bg}$ characteristics at different (e) negetive and (f) positive top gate voltages. Dashed arrows show clockwise hysteresis loops of these characteristics. (g) ON-OFF current ratio and (h) transconductance as a function of top gate voltage. (i) Variation of I$_{ds}$ with V$_{tg}$ at zero back gate voltage. Line arrow shows a counterclockwise I$_{ds}$ loop with top gate voltages. The shaded region shows negetive differential resistance (NDR) state above a critical voltage. (j) Schematic representation of tunneling of charge carriers from channel to CIPS-IPS at some critical voltage.     
\label{Figure4}}
\end{figure}

All the presented evidence unambiguously confirms the existence of FE-PE heterophases in the 2D CIPS-IPS, positioning it as a promising non-volatile element in vdW heterostructures. The material is then tested for its applications in a prototype FEFET. The vdW heterojunction is fabricated by exfoliating thin CIPS-IPS flakes onto WS$_2$ channel material on a Si/SiO$_{2}$ substrate, can be seen in the optical micrograph of Figure \ref{Figure4}a. The experimental configuration of FeFET is depicted in Figure \ref{Figure4}b, where a ferroelectric capacitor is integrated externally with the top gate of WS$_{2}$/SiO$_{2}$. This approach enables the identification of the best match between the negative capacitance (NC) of the ferroelectric and the intrinsic capacitance of the transistor's gate. The equivalent capacitance network can be seen in Figure \ref{Figure4}c.
The NCFET integrates a ferroelectric material into the gate stack of the transistor, demonstrating negative capacitance under specific conditions. Within the NCFET's metal-ferroelectric-insulator-semiconductor (MFIS) configuration, the arrangement can be analogously depicted as a capacitor divider circuit comprising the ferroelectric capacitance, C$_{FE}$, and the inherent baseline FET capacitance, C$_{int}$, as illustrated in Figure \ref{Figure4}c. An inverse C$_{FE}$ yields voltage amplification (A$_{V}$) at the internal gate, where $A_{V}=\frac{\lvert C_{FE}\rvert}{\lvert C_{FE}\rvert-C_{int}}$ \cite{NC_Salahuddin}. The internal voltage amplification plays a crucial role in surpassing the sub-threshold swing beyond the Boltzmann limit. Furthermore, it leads to a higher ON current in NCFET compared to the baseline FET under the same applied voltage. From a thermodynamic perspective, any equilibrium ferroelectric system can be characterized by a double well energy profile at zero applied electric field, as depicted in Figure \ref{Figure4}d. This energy profile features three extremal states: two with positive curvature or positive capacitance minima, and one with negative curvature or a NC maximum, where capacitance equals ($\frac{\partial^2 G}{\partial P^2}-1$) and the energy profile can be understood from the Landau-Khalatnikov (L-K) theory \cite{Amrouch_Neg}. The system typically occupies one of the two states of minimum energy corresponding to a nonzero or remnant polarization at zero bias. The NC state associated with the energy maximum is unstable and thus inaccessible in an isolated ferroelectric capacitor.  
Salahuddin and Datta \cite{NC_Salahuddin} proposed to stabilize the typically unstable NC state by introducing a positive dielectric capacitance in series with the ferroelectric, thereby ensuring a positive total capacitance for the system. By employing this technique, a stable minimum in the total energy of the system (at P$_0$) is achieved, with the ferroelectric residing in the NC state, as depicted in Figure \ref{Figure4}d. In the structure of an NCFET, the internal baseline FET serves as a positive capacitance (C$_{int}$) (Figure \ref{Figure4}d), capable of stabilizing the ferroelectric in the NC state if the total gate capacitance ($\frac{1}{C_{int}} + \frac{1}{C_{FE}})^{-1}$ is positive. This condition necessitates that $\lvert$C$_{FE}$$\rvert$ must exceed C$_{int}$ to realize a thermodynamically stable NC state of the ferroelectric in the NCFET.

Figure \ref{Figure4}e illustrates the input transfer characteristic of an n-type Negative Capacitance NC-FET. The bottom gate voltage (V$_{bg}$) is swept from -10 V to +30 V and back to the starting point while the drain voltage (V$_{sd}$) is set to 3 V. Different curves are plotted with varying top gate voltages (V$_{tg}$) to visualize the impact of it on the transfer characteristics (I$_{sd}$-V$_{bg}$). At zero top gate voltage, a relatively large clockwise hysteresis loop is obtained, typically attributed to the presence of interface traps \cite{kaushik}. Substantial reduction in hysteresis loop in I$_{sd}$-V$_{bg}$ occurs after the application of a small negetive bias at top gate, indicating that the gentle decline characteristic corresponds to the NC effect. Surprisingly, a steeper off-to-on transition of the drain current I$_{sd}$ is observed when the negative top gate voltages are set at and above 5 Volts.
Exactly similar but opposite behavior is obtained on the transfer characteristics for positive V$_{tg}$ as shown in Figure \ref{Figure4}f. The ON/OFF ratio obtained from the transfer characteristics is plotted against V$_{tg}$ and found three different regions, shown in Figure \ref{Figure4}g. At higher positive and negetive V$_{tg}$, after getting sufficient energies the Cu ions starts to migrate through the vdW layers by finding channels through Cu vacancies which remained inactive so far. For higher positive V$_{tg}$, the Cu ions starts to get accumulated near the positively polarized bound charges and the n-channel. Consequently, n-channel carriers get depleted into CIPS-IPS layers via air gap and interfacial gap between WS$_2$ and CIPS-IPS instead of going towards source to drain. For sufficiently large negetive bias V$_{tg}$, Cu ions migrate along gate terminal and get accumulated there and carriers from CIPS-IPS regions get depleted into n-channel. Otherwise, at sufficiently lower V$_{tg}$, when Cu ions cannot participate in long-range conduction, leakage through the FE is negligible and NC effect dominates. At that voltages ($\leq V_{th}$), the ON-OFF ratio follows usual trends (see the zoomed view of ON-OFF ratio in Figure \ref{Figure4}g). The channel transconductance (dI$_{sd}$/{dV$_{bg}$}) against V$_{tg}$ are plotted for different V$_{bg}$ are shown in Figure \ref{Figure4}h. A stable transconductance is obtained for $\leq V_{th}$ which signifies a different state in between two binary state of TFETs. The intermediate logic state in between conventional ON and OFF can be useful to create logic devices with additional logic `1/2' state in between `0' and `1' states. This logic `1/2' state can be obtained by dual gate modulation. We have also measured channel currents as a function of ferroelectric gate voltage (at zero V$_{bg}$).  When the applied voltage (V) exceeds a critical voltage, macroscopic polarization develops into CIPS-IPS, and the device enters the``On" state (see  Figure \ref{Figure4}i ). In this voltage range, intralayer and interlayer Cu ion hopping begins, leading to significant ionic conduction in CIPS-IPS. Consequently, once this critical field is reached, tunneling from the channel to the gate occurs, and the device characteristics enter the region of negative differential resistance (NDR). The tunnelling current direction will get reversed for sufficiently high negative gate bias. This tunnelling phenomena is shown pictorially by 2D schematics (Figure \ref{Figure4}j). 

\section{Conclusion}

In conclusion, negative capacitance Tunnel Field-Effect Transistors (TFETs) based on ferroionic crystals are proposed as one of the most promising devices to generate a stable non-volatile logic state. Experimental validation demonstrates that a less prominent Negative Capacitance (NC) based on a ferroelectric-paraelectric (FE-PE) heterostructure provides beneficial effects on TFET operation and digital performances. This includes an improved off-to-on transition and an extended stable region below the coercive voltage of the NC ferroelectric, facilitated by the tunneling and voltage pinning effects of the ferroelectric material. The study suggests the potential of these devices for stable and efficient non-volatile logic applications. 

\textbf{Supporting Information} Detailed scanning electron microscopy, X-ray diffraction and temperature dependent ferroelectric characterizations supplied as Supporting Information.

%%%%%%%%%%%%%%%%%%%%%%%%%%%%%%%%%%%%
\begin{acknowledgments}
KD would like to acknowledge CSS facilities (XRD, SEM, AFM) of IACS. SD also acknowledges support (lithography facility) from the Technical Research Centre (TRC), IACS, Kolkata. The financial supports from IACS, DST-INSPIRE and CSIR-UGC are greatly acknowledged. SD acknowledges the financial support from DST-SERB grant No. CRG/2021/004334 and special grant SCP/2022/000411.
\newline

S.D., and K.D. conceived the project and designed the experiments. K.D., and B.D. prepared the samples and performed their initial characterization together. K.D., and B.D. performed PFM measurements with the help of S.N. and analyzed the data. Raman measurements and analysis were carried out by K.D. and S. Das. Device fabrication and modification were carried out by T.K., K.D. along with P.K.H., B.K., S.M., P.M., R.P. K.D., B.D., T.K. and R.P. performed the electrical measurements on device and analyzed the data. All authors discussed the results and actively commented on the manuscript written by K.D. and S.D. 
\end{acknowledgments}
%%%%%%%%%%%%%%%%%%%%%%%%%%%%%%%%%%%%

\end{document}